\documentclass[aps,pre,onecolumn,10pt,showpacs,superscriptaddress,groupedaddress,longbibliography]{revtex4-1} 
\pdfoutput=1

\usepackage{graphicx}        
\usepackage{dcolumn}        
\usepackage{bm}             
\usepackage{amssymb}
\usepackage{amsmath}
\usepackage{mathtools}
\usepackage{epstopdf}
\usepackage{color}
\definecolor{korr_26Apr}{rgb}{0,0,0} 
\definecolor{red}{rgb}{1,0,0}

\def \d{\mathrm{d}}

\begin{document}

\widetext

\title{Universal friction law at granular solid-gas transition explains scaling of sediment transport load with excess fluid shear stress}
\author{Thomas P\"ahtz$^{1,2}$}
\email{0012136@zju.edu.cn}
\author{Orencio Dur\'an$^3$}
\affiliation{1.~Institute of Port, Coastal and Offshore Engineering, Ocean College, Zhejiang University, 310058 Hangzhou, China \\
2.~State Key Laboratory of Satellite Ocean Environment Dynamics, Second Institute of Oceanography, 310012 Hangzhou, China \\
3.~Department of Ocean Engineering, Texas A\&M University, College Station, Texas 77843-3136, USA}

\begin{abstract}
A key interest in geomorphology is to predict how the shear stress $\tau$ exerted by a turbulent flow of air or liquid onto an erodible sediment bed affects the transport load $M\tilde g$ (i.e., the submerged weight of transported nonsuspended sediment per unit area) and its average velocity when exceeding the sediment transport threshold $\tau_t$. Most transport rate predictions in the literature are based on the scaling $M\tilde g\propto\tau-\tau_t$, the physical origin of which, however, has remained controversial. Here we test the universality and study the origin of this scaling law using particle-scale simulations of nonsuspended sediment transport driven by a large range of Newtonian fluids. We find that the scaling coefficient is a universal approximate constant and can be understood as an inverse granular friction coefficient (i.e., the ratio between granular shear stress and normal-bed pressure) evaluated at the base of the transport layer (i.e., the effective elevation of energetic particle-bed rebounds). Usually, the granular flow at this base is gaslike and rapidly turns into the solidlike granular bed underneath: a liquidlike regime does not necessarily exist, which is accentuated by a nonlocal granular flow rheology in both the transport layer and bed. Hence, this transition fundamentally differs from the solid-liquid transition (i.e., yielding) in dense granular flows even though both transitions are described by a friction law. Combining this result with recent insights into the nature of $\tau_t$, we conclude that the transport load scaling is a signature of a steady rebound state and unrelated to entrainment of bed sediment.
\end{abstract}

\maketitle

\section{Introduction}
The transport of sediment mediated by the turbulent shearing flow of a Newtonian fluid over an erodible granular bed is responsible for the evolution of fluid-sheared surfaces composed of loose sediment, such as river and ocean beds, and wind-blown sand surfaces on Earth and other planets, provided that the sediment is not kept suspended by the fluid turbulence \cite{Bagnold41,Yalin77,Graf84,vanRijn93,Julien98,Garcia07,Bourkeetal10,PyeTsoar09,Zheng09,Shao08,Duranetal11,Koketal12,Rasmussenetal15,Valanceetal15}. Nonsuspended sediment transport thus constitutes one of the most important geomorphological processes in which granular particles collectively move like a continuum flow, and predicting the associated sediment transport rate $Q$ (i.e., the total particle momentum in the flow direction per unit bed area) and flow threshold $\tau_t$ (i.e., the value of the fluid shear stress $\tau$ below which sediment transport ceases) are considered central problems in Earth and planetary geomorphology \cite{Bagnold41,Yalin77,Graf84,vanRijn93,Julien98,Garcia07,Bourkeetal10,PyeTsoar09,Shao08,Zheng09,Duranetal11,Koketal12,Rasmussenetal15,Valanceetal15}. Here we provide the theoretical base necessary to understand the scaling of $Q$ and $\tau_t$ and, by doing so, show that and why nonsuspended sediment transport constitutes a class of granular flows with unique properties, such as a nonlocal granular flow rheology even relatively far from the flow threshold.

\subsection{The scaling of the transport rate of nonsuspended sediment}
\begin{figure}[htb]
 \begin{center}
  \includegraphics[width=0.5\columnwidth]{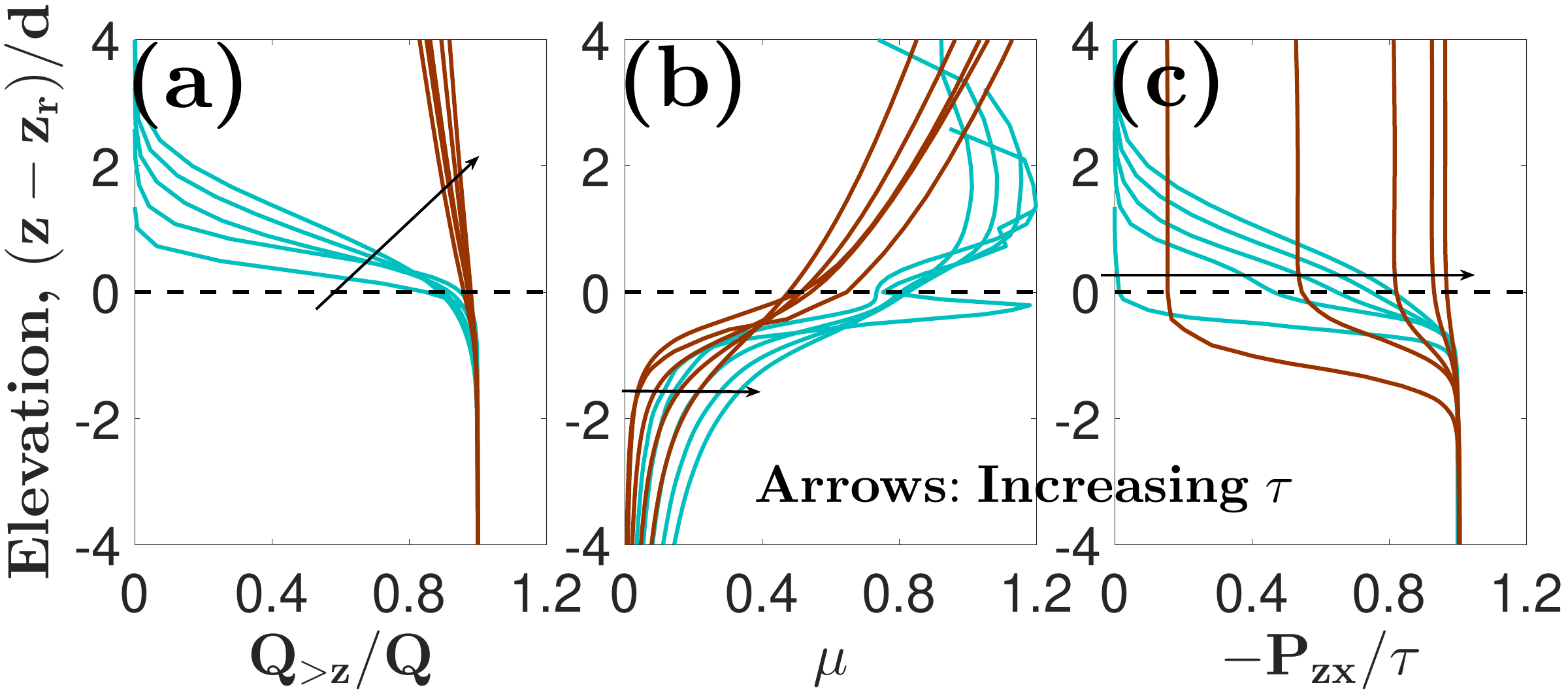}
 \end{center}
 \caption{{\bf{Visualization of Bagnold interface properties.}} Vertical profiles of (a) the fraction $Q_{>z}/Q$ of sediment transport occurring above elevation $z$, (b) the friction coefficient $\mu$, and (c) the ratio $-P_{zx}/\tau$ between the particle shear stress $-P_{zx}$ and fluid shear stress $\tau$. The solid lines correspond to data obtained from direct sediment transport simulations (see Sec.~\ref{NumericalSimulations}) for two representative cases: turbulent bedload (turquoise) and saltation transport (brown). The black, dashed lines mark the Bagnold interface $z=z_r$.}
 \label{SketchBagnoldInterface}
\end{figure}

Numerous experimental and theoretical studies (e.g., Refs. \cite{MeyerPeterMuller48,Einstein50,Yalin63,Bagnold56,Bagnold66,Bagnold73,AshidaMichiue72,EngelundFredsoe76,KovacsParker94,NinoGarcia94,NinoGarcia98a,Seminaraetal02,Parkeretal03,AbrahamsGao06,FernandezLuqueBeek76,Smart84,Lajeunesseetal10,CapartFracarollo11,DoorschotLehning02,HanesBowen85,Ninoetal94,NinoGarcia98b,CharruMouilleronArnould02,Charru06,Berzi11,Berzi13,Charruetal16,Maurinetal18,UngarHaff87,Almeidaetal06,Almeidaetal07,Almeidaetal08,Reckingetal08b,Creysselsetal09,Hoetal11,MartinKok17,Duranetal12,Aussillousetal13,AliDey17,Kawamura51,Owen64,Kind76,LettauLettau78,Sorensen91,Sorensen04,Sauermannetal01,DuranHerrmann06,Pahtzetal12,Lammeletal12,JenkinsValance14,Berzietal16,Huangetal14,WangZheng15}) have measured or derived analytical expressions for the transport rate $Q$ as a function of particle and environmental parameters, such as the particle (fluid) density $\rho_p$ ($\rho_f$), kinematic fluid viscosity $\nu$, characteristic particle diameter $d$, gravitational constant $g$, and $\tau$ and $\tau_t$. Most of the theoretical derivations are based on, or can be reformulated in the spirit of, Bagnold's \cite{Bagnold56,Bagnold66,Bagnold73} pioneering ideas. Defining a Cartesian coordinate system $\mathbf{x}=(x,y,z)$, where $x$ is in the flow direction, $z$ in the direction normal to the bed oriented upwards, and $y$ in the lateral direction, Bagnold assumed that there is a well-defined interface $z=z_r$ between granular bed ($z<z_r$) and transport layer ($z>z_r$), which we henceforth call the ``Bagnold interface,'' with the following properties (Fig.~\ref{SketchBagnoldInterface}):
\begin{enumerate}
 \item The transport rate $Q_r$ above $z_r$ well approximates the total transport rate $Q$ (i.e., $z_r$ cannot be too far away from the actual granular bed). Hence, one can separate $Q$ into the mass $M=\rho_p\int_{z_r}^\infty\phi\d z$ of particles located above $z_r$ per unit bed area, where $\phi$ is the particle volume fraction (i.e., the fraction of space covered by particles), and the average horizontal velocity $\overline{v_x}$ with which particles located above $z_r$ move: $\overline{v_x}\equiv Q_r/M\simeq Q/M$. \label{Property1}
 \item The ratio $\mu\equiv-P_{zx}/P_{zz}$ between the particle shear stress $-P_{zx}$ and normal-bed pressure $P_{zz}$, where $P_{ij}$ is the particle stress tensor, at $z_r$ does not significantly depend on the fluid shear stress $\tau$: $\mu_b\equiv\mu(z_r)\ne f(\tau)$. \label{Property2}
 \item The ratio $-P_{zx}(z_r)/\tau$ between particle and fluid shear stress increases from nearly zero at low transport stages ($\tau/\tau_t-1\ll1$) to nearly unity at large transport stages ($\tau/\tau_t-1\gg1$). Two simple expressions that obey this constraint are $-P_{zx}(z_r)=\tau-\tau_t$ and $-P_{zx}(z_r)=\sqrt{\tau}(\sqrt{\tau}-\sqrt{\tau_t}$). Note that the former expression is usually attributed to Owen \cite{Owen64} (``Owen's second hypothesis'' \cite{Walteretal14}) in the aeolian transport literature \cite{Shao08,Duranetal11,Koketal12} even though Bagnold \cite{Bagnold56} was its originator and also applied it to aeolian transport. \label{Property3}
\end{enumerate}
Combining these three properties and using the vertical momentum balance $P^\prime_{zz}\simeq-\rho_p\phi\tilde g$ of steady, homogeneous sediment transport \cite{Pahtzetal15a}, where the prime denotes the derivative $\d/\d z$ and $\tilde g=(1-\rho_f/\rho_p)g$ the buoyancy-reduced value of $g$, then yields
\begin{equation}
   \begin{matrix*}[l]
   Q\simeq\mu_b^{-1}\tilde g^{-1}(\tau-\tau_t)\overline{v_x} & \;\text{if} & \;-P_{zx}(z_r)=\tau-\tau_t, \\
	 Q\simeq\mu_b^{-1}\tilde g^{-1}\sqrt{\tau}(\sqrt{\tau}-\sqrt{\tau_t})\overline{v_x} & \;\text{if} & \;-P_{zx}(z_r)=\sqrt{\tau}(\sqrt{\tau}-\sqrt{\tau_t}).
 \end{matrix*}
 \label{Q}
\end{equation}
Indeed, the functional behaviors in Eq.~(\ref{Q}) resemble the vast majority of theoretical and experimental threshold shear stress-based expressions for the transport load $M\tilde g\simeq Q\tilde g/\overline{v_x}$ and transport rate $Q$ in the literature, which differ only in their prediction of $\overline{v_x}$. For example, experiments of nonsuspended sediment transport driven by turbulent streams of liquid (turbulent ``bedload'') suggest that $\overline{v_x}$ is linear in $\sqrt{\tau/\rho_f}$ \cite{FernandezLuqueBeek76,Smart84,Lajeunesseetal10,CapartFracarollo11}, whereas experiments of nonsuspended sediment transport driven by turbulent streams of air (turbulent ``saltation'') suggest that $\overline{v_x}$ is constant with $\tau$ \cite{Creysselsetal09,Hoetal11,MartinKok17}. The capability of Eq.~(\ref{Q}) to reproduce experimental data is indirect evidence that the Bagnold interface exists for these conditions. However, there are a number of unsolved problems, even inconsistencies, regarding the generality and physical origin of the Bagnold interface that currently prevent us from understanding and predicting the scaling laws of nonsuspended sediment transport for arbitrary conditions and from integrating nonsuspended sediment transport within the framework of granular flow rheology.

\subsection{Open questions}
\subsubsection{Existence of the Bagnold interface} \label{Existence}
Natural granular beds are locally very heterogeneous and undergo continuous rearrangements during sediment transport, which renders the definition of a bed-transport-layer interface difficult. For steady, homogeneous transport conditions, four different definitions have been proposed in the literature: the elevation at which the friction coefficient $\mu$ exhibits a certain constant value \cite{Maurinetal15}, the elevation at which the particle volume fraction $\phi$ exhibits a certain constant portion of the bed packing fraction $\phi_b$ \cite{Duranetal12}, the elevation at which the particle shear rate $\dot\gamma$ exhibits a certain constant portion of its maximal value \cite{CapartFracarollo11}, and the elevation at which the production rate $P_{zz}\dot\gamma$ of cross-correlation fluctuation energy is maximal \cite{PahtzDuran17,PahtzDuran18a}. However, whether any of these interfaces is the Bagnold interface and whether the Bagnold interface even exists for nonsuspended sediment transport in arbitrary environments remain unclear.

In this study, we provide answers to the following questions:
\begin{itemize}
 \item Does the Bagnold interface exist in general settings?
 \item If so, is there a general definition of the Bagnold interface?
\end{itemize}

\subsubsection{Physical origin of friction law} \label{CoulombFriction}
Property~\ref{Property2} of the Bagnold interface represents a macroscopic, dynamic friction law, analogous to Coulomb friction describing the sliding of an object down an inclined plane, where the constant dynamic bed friction coefficient $\mu_b$ is the analog to the ratio between the horizontal and normal force acting on the sliding object. In the context of dense ($\phi\gtrsim0.4$) granular flows and suspensions, it is well established that a constant dynamic friction coefficient (the yield stress ratio) characterizes the transition between solidlike and liquidlike flow behavior \cite{CourrechduPontetal03,MiDi04,Cassaretal05,Jopetal06,ForterrePouliquen08,Andreottietal13,Jop15,Boyeretal11,Trulssonetal12,NessSun15,NessSun16,Amarsidetal17,Maurinetal16,Houssaisetal16,HoussaisJerolmack17,Delannayetal17,Royetal17,KamrinKoval12,Bouzidetal13,Bouzidetal15}. Here liquidlike behavior refers to dense flows that obey a local rheology (i.e., $\mu$ depends only on a single local quantity, such as $\phi$), while solidlike behavior refers to both quasistatic and creeping flows (not to be confused with Bagnold's term ``surface creep'' \cite{Bagnold41}). Quasistatic flows are associated with very small, reversible deformations of dense packed granular systems, while creeping flows are associated with an exponential relaxation of the particle shear rate $\dot\gamma$ between quasistatic and liquidlike flows \cite{Nicholetal10,Reddyetal11,KamrinKoval12,Bouzidetal13,Bouzidetal15,Houssaisetal15,AllenKudrolli18} and characterized by a nonlocal granular flow rheology \cite{KamrinKoval12,Bouzidetal13,Bouzidetal15}. Based on the fact that a friction law characterizes the solid-liquid transition, it has been very common to argue that the Bagnold interface separates a solidlike granular bed from a liquidlike transport layer on its top and that $\mu_b$ is the yield stress ratio \cite{AshidaMichiue72,EngelundFredsoe76,KovacsParker94,NinoGarcia94,NinoGarcia98a,Seminaraetal02,Parkeretal03,AbrahamsGao06}, which is in the spirit of Bagnold's original reasoning \cite{Bagnold56,Bagnold66,Bagnold73}. However, this interpretation is inconsistent with Property~\ref{Property3} of the Bagnold interface, which predicts that the particle shear stress $-P_{zx}(z_r)$, and thus the particle volume fraction $\phi(z_r)$ \cite{Pahtzetal15a}, becomes very small when the fluid shear stress approaches the flow threshold ($\tau\rightarrow\tau_t$). It is further inconsistent with the fact that the Bagnold interface is also found in highly simplified numerical sediment transport simulations that do not resolve particle interactions \cite{NinoGarcia98a,Pahtzetal12}.

An alternative interpretation of the friction law came from studies on saltation transport \cite{Sauermannetal01,DuranHerrmann06,Pahtzetal12,Lammeletal12,JenkinsValance14,Berzietal16,Berzietal17}. They suggested that $\mu_b$ is an effective restitution coefficient characterizing an approximately constant ratio between the average horizontal momentum loss and vertical momentum gain of particles rebounding at the Bagnold interface. However, this interpretation has never been tested against experiments or numerical particle-scale simulations of sediment transport, and it is unclear how it can be generalized to the bedload transport regime, in which transported particles experience long-lasting contacts with the granular bed and each other \cite{Schmeeckle14}.

In this study, we provide answers to the following questions:
\begin{itemize}
 \item What is the physical origin of the friction law at the Bagnold interface?
 \item Is this origin in some way associated with the rheology of dense granular flows and suspensions?
\end{itemize}

\subsubsection{Universality of friction law} \label{Universality}
For the purpose of understanding the scaling laws of nonsuspended sediment transport in arbitrary environments, it is crucial to know how much the dynamic bed friction coefficient $\mu_b$ at the Bagnold interface varies with environmental parameters other than $\tau$. Currently, the literature suggests that the friction coefficient $\mu$ at elevations near the bed surface, and thus near the Bagnold interface, strongly depends on the fluid driving transport (reported values range from $0.2$ in water \cite{NinoGarcia98b} to $1.0$ in air \cite{Pahtzetal12}), which if true would imply that the friction law is not universal. However, particle stresses are notoriously difficult to measure in erodible granular beds \cite{Delannayetal17}, which is why either measurements of $\mu$ have been limited to systems that only crudely represent natural nonsuspended sediment transport, such as the motion of externally fed particles along rigid beds \cite{Francis73,AbbottFrancis77,NinoGarcia98b}, or $\mu$ has been estimated as $\tau/P_{zz}$ \cite{HanesInman85}, which makes sense only for intense transport conditions due to Property~\ref{Property3}.

In this study we provide an answer to the following question:
\begin{itemize}
 \item How much does the dynamic friction coefficient $\mu_b$ at the Bagnold interface vary with environmental parameters?
\end{itemize}

\subsection{Organization of this paper}
The method that we use to answer the open questions outlined above, direct sediment transport simulations with the model of Ref.~\cite{Duranetal12}, is briefly introduced in Sec.~\ref{NumericalSimulations}. Section~\ref{CoulombFrictionLaw} then puts forward our definition of the bed-transport-layer interface as the effective elevation at which the most energetic transported particles rebound when colliding with bed surface particles and shows that this interface is the Bagnold interface. It also shows that the friction law at the Bagnold interface is, indeed, universal. Section~\ref{PhysicalOrigin} links this finding, for the vast majority of sediment transport regimes, to a steady transport state in which transported particles continuously rebound at the bed surface and shows that alternative explanations associated with the rheology of dense granular flows and suspensions in general fail due to the absence of a liquidlike flow regime. Finally, Sec.~\ref{Conclusions} summarizes the main conclusions that can be drawn from our results and discusses our results in the context of sediment transport modeling.

\section{Numerical simulations} \label{NumericalSimulations}
In this section, we describe the numerical model (Sec.~\ref{NumericalModel}), the simulated sediment transport conditions (Sec.~\ref{SimulatedConditions}), and how we use the simulation data to compute relevant physical quantities (Sec.~\ref{Computation}).
\subsection{Numerical model description} \label{NumericalModel}
The numerical model of sediment transport in a Newtonian fluid of Ref.~\cite{Duranetal12} belongs to a new generation of sophisticated grain-scale models of sediment transport \cite{Carneiroetal11,Duranetal11,Duranetal12,Carneiroetal13,Jietal13,Duranetal14a,Duranetal14b,KidanemariamUhlmann14a,KidanemariamUhlmann14b,KidanemariamUhlmann17,Schmeeckle14,Vowinckeletal14,Vowinckeletal16,ArollaDesjardins15,Pahtzetal15a,Pahtzetal15b,Carneiroetal15,Clarketal15,Clarketal17,Derksen15,Maurinetal15,Maurinetal16,FinnLi16,Finnetal16,SunXiao16,ElghannayTafti17a,ElghannayTafti17b,Gonzalezetal17,Chengetal18,Seiletal18,PahtzDuran17,PahtzDuran18a} and has been shown to reproduce many observations concerning viscous and turbulent nonsuspended sediment transport in air and water \cite{Duranetal11,Duranetal12,Duranetal14a,PahtzDuran17,PahtzDuran18a}, and bedform formation \cite{Duranetal14b}. It couples a discrete element method for the particle motion with a continuum Reynolds-averaged description of hydrodynamics, which means that it neglects turbulent fluctuations around the mean turbulent flow. It simulates the translational and rotational dynamics of $\approx15,000$ spheres, including $>10$ layers of bed particles (more than sufficient to completely dissipate the energy of particles impacting the bed surface), with diameters $d_p$ evenly distributed within two sizes ($0.8d$ and $1.2d$) in a quasi-2-D, vertically infinite domain of length $1181d$. Periodic boundary conditions are imposed along the flow direction, while the bottommost layer of particles is glued to a bottom wall. The particle contact model considers normal repulsion (restitution coefficient $e$), energy dissipation, and tangential friction, where the magnitude of the tangential friction force relative to the normal contact force is limited through a Coulomb friction criterion (contact friction coefficient $\mu_c=0.5$). The Reynolds-averaged Navier-Stokes equations are applied to an inner turbulent boundary layer of infinite size, which means that the flow depth of fluvial flows is assumed to be much larger than the thickness of the bedload transport layer. These equations are combined with an improved mixing length approximation that ensures a smooth hydrodynamic transition from high to low particle concentration at the bed surface and quantitatively reproduces the law of the wall flow velocity profile in the absence of transport. The model considers the gravity, buoyancy, added-mass, and fluid drag force acting on particles. However, cohesive and higher-order fluid forces, such as the lift force and hindrance effect on the drag force are neglected, while lubrication forces are considered indirectly through varying $e$ (Sec.~\ref{SimulatedConditions}). We refer the reader to the original publication \cite{Duranetal12} for further details (note that we recently corrected slight inaccuracies in the original model \cite{PahtzDuran17}).

\subsection{Simulated sediment transport conditions} \label{SimulatedConditions}
Using the numerical model, we simulate steady, homogeneous sediment transport for a particle-fluid-density ratio $s\equiv\rho_p/\rho_f$ within the range $s\in[1.1,2000]$, a Galileo number $\mathrm{Ga}\equiv\sqrt{(s-1)gd^3}/\nu$ within the range $\mathrm{Ga}\in[0.1,100]$, and a normal restitution coefficient of dry binary collisions of $e=0.9$. For small density ratio ($s\leq2.65$), we also carry out simulations with $e=0.01$ because $e$ can become very small for small Stokes numbers due to lubrication forces \cite{Gondretetal02,YangHunt06,SimeonovCalantoni12}. For each set of $s$, $\mathrm{Ga}$, and $e$, we vary the dimensionless fluid shear stress (``Shields number'') $\Theta=\tau/[(\rho_p-\rho_f)gd]$ in regular intervals above its threshold value $\Theta_t=\tau_t/[(\rho_p-\rho_f)gd]$, which we obtain from extrapolation to vanishing transport \cite{PahtzDuran18a}. The simulated conditions cover four major, and very distinct, natural transport regimes, which depend on the transport layer thickness and the thickness of the viscous sublayer of the turbulent boundary layer \cite{PahtzDuran18a}: viscous bedload transport, such as the transport of sand by oil; turbulent bedload transport, such as the transport of gravel by water; viscous saltation transport, such as the transport of sand by wind on Mars; and turbulent saltation transport, such as the transport of sand by wind on Earth. They also cover $5$ orders of magnitude of the `impact number' $\mathrm{Im}\equiv\sqrt{s+0.5}\mathrm{Ga}\simeq\sqrt{s}\mathrm{Ga}$, which characterizes the mode of entrainment of bed sediment under threshold conditions \cite{PahtzDuran17}: $\mathrm{Im}\gtrsim20$ when entrainment by particle-bed impacts dominates entrainment by the mean turbulent flow, $\mathrm{Im}\lesssim5$ when direct entrainment by the mean turbulent flow dominates, and transitional behavior when $5\lesssim\mathrm{Im}\lesssim20$.

\subsection{Computation of local averages and particle stresses} \label{Computation}
We use the simulation data to compute local averages of particle properties and the particle stress tensor, which is explained in the following.

\subsubsection{Local, mass-weighted time average and particle volume fraction}
We compute the local, mass-weighted time average $\langle A\rangle$ of a particle quantity $A$ through \cite{Pahtzetal15a}
\begin{eqnarray}
 \langle A\rangle&=&\frac{1}{\Delta\phi}\overline{\sum_nV_p^nA^n\delta(z-z^n)}^T, \label{LocalAverage} \\
 \phi&=&\frac{1}{\Delta}\overline{\sum_nV_p^n\delta(z-z^n)}^T, 
\end{eqnarray}
where $\Delta=1181d^2$ is the simulation area, $\phi$ is the local particle volume fraction, $z^n$ ($V^n_p=\pi d_p^{n3}/6$) is the elevation (volume) of particle $n$, $\delta$ the $\delta$ distribution, and $\overline{\cdot}^T=\frac{1}{T}\int_0^T\cdot\d t$ denotes the time average over a sufficiently long time $T$. The $\delta$ kernels have been coarse grained through spatial averaging over a discretization box of size $1181d\times d\times\Delta z$, where $\Delta z$ varies between $0.05d$ in dense and dilute flow regions ($\phi\gtrsim0.1$) and larger values in rarefied regions. Henceforth, the $\delta$ symbol should thus be interpreted as the associated coarse-graining function.

\subsubsection{Particle stress tensor}
The particle stress tensor $P_{ij}$ is composed of a kinetic contribution due to the transport of momentum between contacts (superscript `t') and a contact contribution (superscript `c') and computed through \cite{Pahtzetal15a}
\begin{subequations}
\begin{eqnarray}
 P_{ij}&=&P^t_{ij}+P^c_{ij}, \label{P} \\
 P^t_{ij}&=&\rho_p\phi\langle c_ic_j\rangle, \label{Pt} \\
 P^c_{ij}&=&\frac{1}{2\Delta}\overline{\sum_{mn}F_j^{mn}(x^m_i-x^n_i)K(z,z^m,z^n)}^T,
\end{eqnarray}
\end{subequations}
where $K=\int\limits_0^1\delta\{z-[(z^m-z^n)\tilde s+z^n]\}\d\tilde s$, $\mathbf{c}=\mathbf{v}-\langle\mathbf{v}\rangle$ is the fluctuation velocity, and $\mathbf{F}^{mn}$ the contact force applied by particle $n$ on particle $m$ ($\mathbf{F}^{mm}=0$). We confirmed that these definitions are consistent with the steady momentum balance $P^\prime_{zi}=\rho_p\phi\langle a_i\rangle$ \cite{Pahtzetal15a}, where $\mathbf{a}$ is the particle acceleration due to noncontact forces.

\section{Existence of the Bagnold interface in arbitrary environments} \label{CoulombFrictionLaw}
In Sec.~\ref{Interface}, we first put forward our definition of the bed-transport-layer interface. In Sec.~\ref{TestInterface}, we then show with data from our direct transport simulations that this definition, in contrast to common alternative definitions, obeys the properties of the Bagnold interface (except for a slight restriction regarding Property~\ref{Property3}) with a universally approximately constant bed friction coefficient $\mu_b$.

\subsection{Definition of the bed-transport-layer interface} \label{Interface}
In order to motivate a definition of the bed-transport-layer interface that results in the Bagnold interface, we exploit the fact that numerical studies that represent the granular bed surface by a rigid bottom wall found that this wall obeys Properties~\ref{Property1}-\ref{Property3} of the Bagnold interface \cite{NinoGarcia98b,Pahtzetal12}. This finding suggests that an appropriate definition should have characteristics that mimic those of particle rebounds at rigid boundaries. One such characteristic is the production of particle velocity fluctuations. For example, gravity-driven granular flows down an inclined, rigid base exhibit a maximum of the granular temperature $\langle\mathbf{c}^2\rangle$ near this base \cite{Broduetal15}. The probable reason is that such rigid boundaries induce strong correlations between the velocities of descending particles before rebound and ascending particles after rebound.

In steady sediment transport, the mass balance dictates $\langle v_z\rangle=0$ \cite{Pahtzetal15a}, which can be achieved only if rebounds of transported particles at the granular bed partially convert horizontal momentum of descending particles into vertical momentum of ascending particles (i.e., negative correlation). Similar to gravity-driven granular flows, this constraint implies that particle-bed rebounds are a strong source of the negative cross-correlation fluctuation energy density $-\rho_p\phi\langle c_zc_x\rangle$.

\begin{figure}[htb]
 \begin{center}
  \includegraphics[width=1.0\columnwidth]{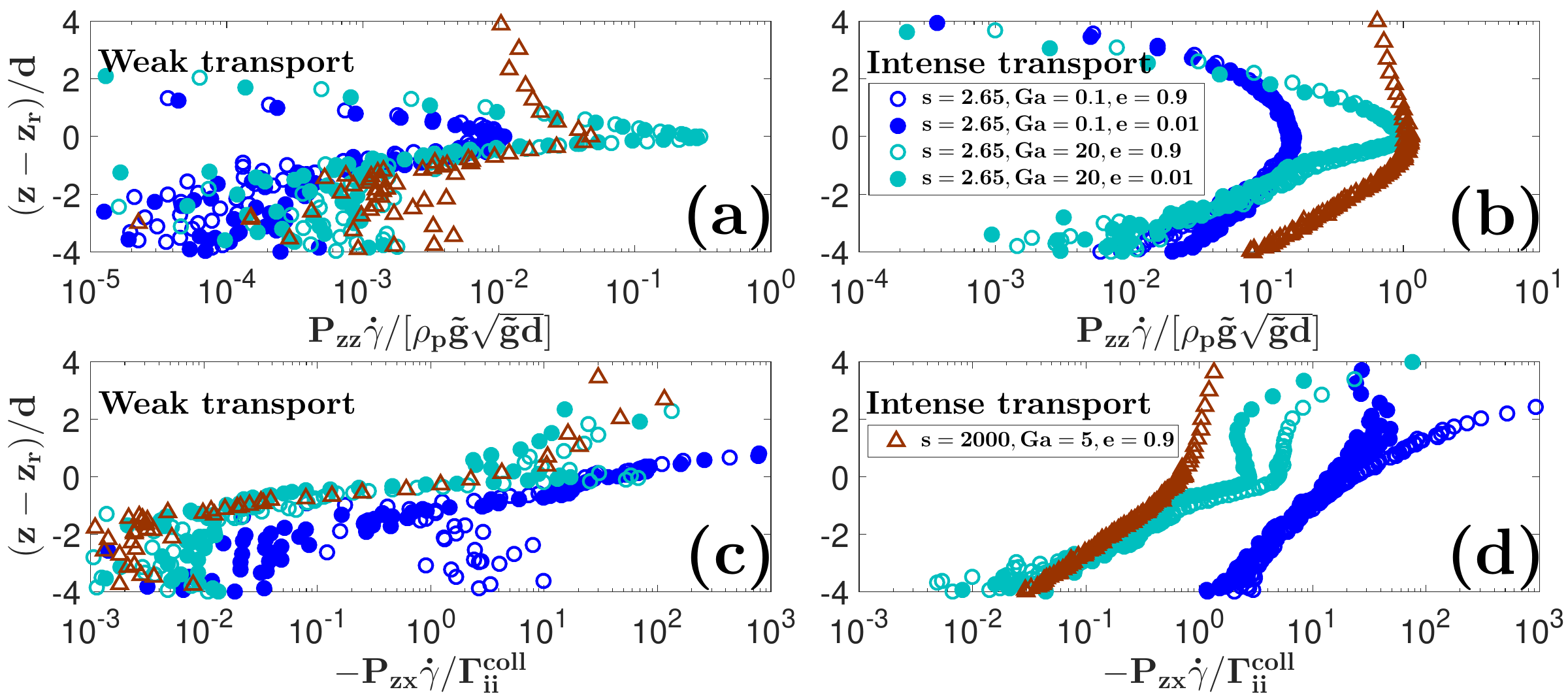}
 \end{center}
 \caption{{\bf{Exemplary vertical profiles of quantities associated with the (cross-correlation) fluctuation energy balance.}} (a, b) Vertical profiles relative to the rebound location $z_r$ of $P_{zz}\dot\gamma/(\rho_p\tilde g\sqrt{\tilde gd})$. (c, d) Vertical profiles of $-P_{zx}\dot\gamma/\Gamma^{\mathrm{coll}}_{ii}$. Symbols correspond to viscous bedload transport [$s=2.65$, $\mathrm{Ga}=0.1$, $\Theta/\Theta_t=(1.9,6.2)$], turbulent bedload transport [$s=2.65$, $\mathrm{Ga}=20$, $\Theta/\Theta_t=(2.0,7.2)$], and turbulent saltation transport [$s=2000$, $\mathrm{Ga}=5$, $\Theta/\Theta_t=(2.3,49)$], where the data with smaller values of the rescaled Shields number $\Theta/\Theta_t$ are shown in (a) and (c) and those with larger values of $\Theta/\Theta_t$ in (b) and (d). For the bedload transport conditions, the restitution coefficient has been varied to mimic the minimal ($e=0.9$) and nearly maximal ($e=0.01$) possible effect of lubrication forces.}
 \label{Profiles}
\end{figure}

The balances of $-\rho_p\phi\langle c_zc_x\rangle$ and of the actual fluctuation energy density $\rho_p\phi\langle\mathbf{c}^2\rangle$ can be derived rigorously from Newton's axioms. For steady sediment transport ($\partial/\partial x=\partial/\partial y=\partial/\partial t=0$), they read \cite{Pahtzetal15a} (Einsteinian summation)
\begin{subequations}
\begin{eqnarray}
 -q^\prime_{z(xz)}&=&\frac{1}{2}P_{zz}\dot\gamma+\Gamma^{\mathrm{drag}}_{(xz)}+\Gamma^{\mathrm{coll}}_{(xz)}, \label{CrossFlucEnergy} \\
  q^\prime_{zii}&=&-P_{zx}\dot\gamma-\Gamma^{\mathrm{drag}}_{ii}-\Gamma^{\mathrm{coll}}_{ii}, \label{FlucEnergy}
\end{eqnarray}
\end{subequations}
respectively, where the parentheses denote the symmetrization in the indices [$A_{(ij)}=\frac{1}{2}(A_{ij}+A_{ji})$]. Furthermore, $q_{ijk}=\frac{\rho_p\phi}{2}\langle c_ic_jc_k\rangle+\frac{1}{2\Delta}\overline{\sum_{mn}F^{mn}_jc_k(x^m_i-x^n_i)K(z,z^m,z^n)}^T$ is the flux tensor of fluctuation energy, $\dot\gamma=\langle v_x\rangle^\prime$ the particle shear rate, $\Gamma^{\mathrm{drag}}_{ij}=-\rho_p\phi\langle a_ic_j\rangle$ the drag dissipation rate tensor, and $\Gamma^{\mathrm{coll}}_{ij}=-\frac{1}{2\Delta}\overline{\sum_{mn}F^{mn}_i(v^m_j-v^n_j)\delta(z-z^m)}^T$ the collisional dissipation rate tensor. In Eq.~(\ref{FlucEnergy}), $-P_{zx}\dot\gamma$ corresponds to the production rate and $\Gamma^{\mathrm{drag}}_{ii}$ and $\Gamma^{\mathrm{coll}}_{ii}$ to the dissipation rate of $\rho_p\phi\langle\mathbf{c}^2\rangle$ by fluid drag and collisions, respectively. In Eq.~(\ref{CrossFlucEnergy}), $\frac{1}{2}P_{zz}\dot\gamma$ corresponds to the production rate and $-\Gamma^{\mathrm{drag}}_{(xz)}$ and $-\Gamma^{\mathrm{coll}}_{(xz)}$ to the dissipation rate of $-\rho_p\phi\langle c_zc_x\rangle$ by fluid drag and collisions, respectively. Hence, if we identify the bed-transport-layer interface as the average elevation of energetic particle-bed rebounds and use that such rebounds are a strong source of $-\rho_p\phi\langle c_zc_x\rangle$, it makes sense to define this interface through a maximum of the local production rate of $-\rho_p\phi\langle c_zc_x\rangle$:
\begin{equation}
 \max(P_{zz}\dot\gamma)=[P_{zz}\dot\gamma](z_r), \label{InterfaceDef}
\end{equation}
which is exactly the definition that we applied in two recent studies \cite{PahtzDuran17,PahtzDuran18a}. Figures~\ref{Profiles}(a) and \ref{Profiles}(b) show exemplary vertical profiles relative to $z_r$ of $P_{zz}\dot\gamma$ for (a) weak and (b) intense viscous and turbulent bedload transport and turbulent saltation transport, where the bedload cases have been simulated using two different restitution coefficients to mimic the minimal ($e=0.9$) and nearly maximal ($e=0.01$) effect that lubrication forces can possibly have.
It can be seen that the value of $e$ does not significantly affect these profiles. As we will see later, the influence of $e$ on bedload transport properties is very small in general, consistent with previous studies \cite{DrakeCalantoni01,Maurinetal15,ElghannayTafti17b,PahtzDuran17,PahtzDuran18a}.

The interface $z=z_r$ defined by Eq.~(\ref{InterfaceDef}) shares some similarities with the region in which the production rate of fluctuation energy is nearly balanced by the collisional energy dissipation rate: $-P_{zx}\dot\gamma\simeq\Gamma^{\mathrm{coll}}_{ii}$. For turbulent bedload transport, it has been speculated that this region is a distinct granular layer (the ``dense algebraic layer'') with a thickness of several particle diameters $d$ and that the bottom of this layer corresponds to the bed-transport-layer interface \cite{Berzi11,Berzi13}. However, Figs.~\ref{Profiles}(c) and \ref{Profiles}(d) show for the same cases as before that the thickness of the region in which $-P_{zx}\dot\gamma/\Gamma^{\mathrm{coll}}_{ii}\simeq1$ is usually very small ($\ll d$), especially for bedload transport, regardless of whether transport is weak or intense. In order words, the dense algebraic layer usually does not exist. One of the reasons may be the fact that drag dissipation ($\Gamma^{\mathrm{drag}}_{ii}$), which has been neglected in Refs.~\cite{Berzi11,Berzi13}, actually dominates collisional dissipation ($\Gamma^{\mathrm{coll}}_{ii}$) in bedload transport [Fig.~1(b) in Ref.~\cite{Pahtzetal15a}, which is based on the same numerical model].

\subsection{Test of interface definition against data from our direct transport simulations} \label{TestInterface}
\begin{figure}[htb]
 \begin{center}
  \includegraphics[width=1.0\columnwidth]{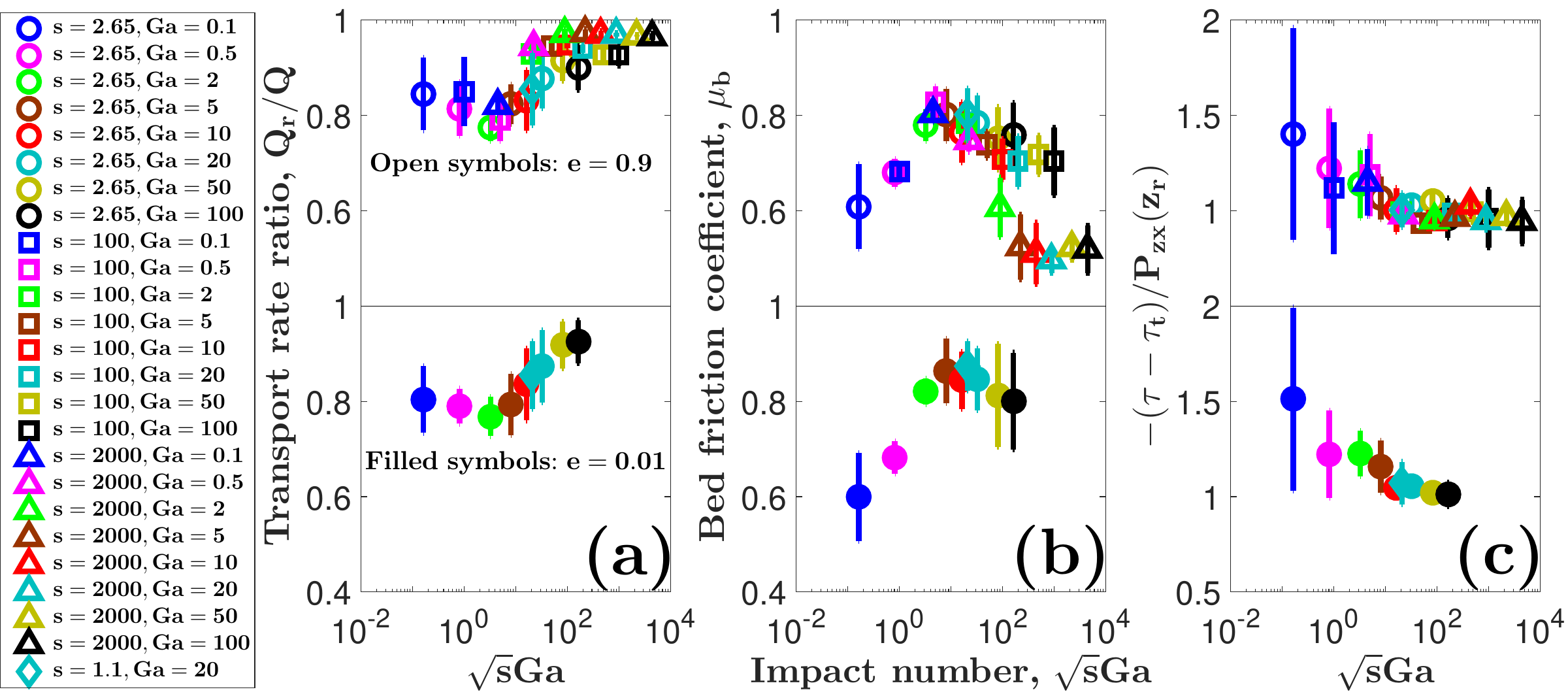}
 \end{center}
 \caption{{\bf{Test of Bagnold interface properties.}} Test of (a) Property~\ref{Property1}, (b) Property~\ref{Property2}, and (c) Property~\ref{Property3} of the Bagnold interface with data from our direct transport simulations for various combinations of the particle-fluid-density ratio $s$, Galileo number $\mathrm{Ga}$, Shields number $\Theta$, and thus impact number $\sqrt{s}\mathrm{Ga}$. For conditions with $s\leq2.65$ (corresponding to bedload transport), the restitution coefficient has been varied to mimic the minimal ($e=0.9$) and nearly maximal ($e=0.01$) possible effect of lubrication forces. The vertical bars indicate the range of values the quantities cover with varying $\Theta$ above about $2\Theta_t$. This lower limit is imposed to separate the random variability due to bad statistics when $\Theta$ is close to $\Theta_t$ [e.g., see Fig.~\ref{BagnoldInterfaceTrends}(c)] from the actual variability. Indications that the Bagnold interface properties are obeyed: (a) the sediment transport rate ratio $Q_r/Q$ is near unity, (b) the bed friction coefficient $\mu_b$ is approximately constant with $\Theta$ (relatively small vertical bars), and (c) the quantity $-(\tau-\tau_t)/P_{zx}(z_r)$ is near unity.}
 \label{BagnoldInterface}
\end{figure}

\begin{figure}[htb]
 \begin{center}
  \includegraphics[width=1.0\columnwidth]{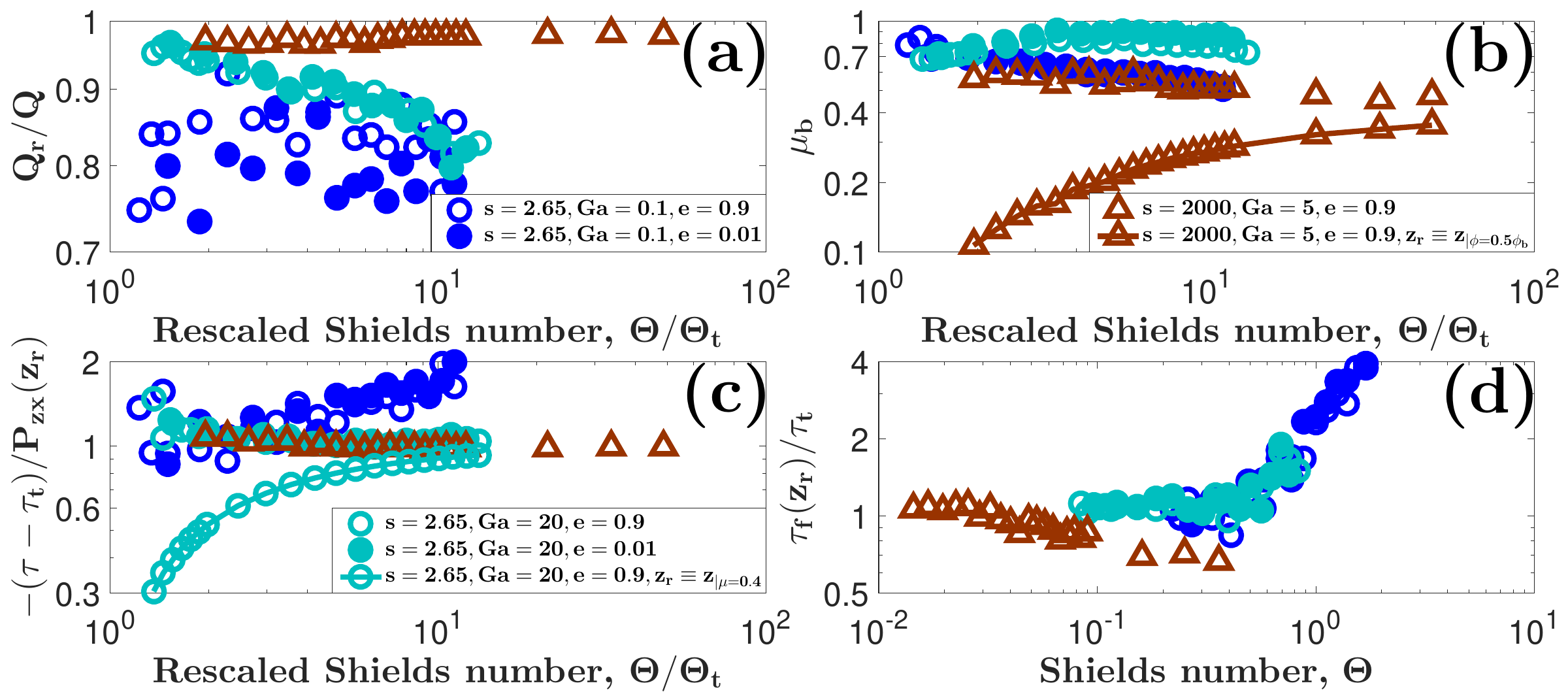}
 \end{center}
 \caption{{\bf{Exemplary trends of quantities associated with Bagnold interface properties.}} (a) Sediment transport rate ratio $Q_r/Q$, (b) bed friction coefficient $\mu_b$, and (c) $-(\tau-\tau_t)/P_{zx}(z_r)$ versus rescaled Shields number $\Theta/\Theta_t$. (d) Rescaled surface fluid shear stress $\tau_f(z_r)/\tau_t$ versus Shields number $\Theta$. The interface $z=z_r$ is calculated by Eq.~(\ref{InterfaceDef}) if not otherwise stated in the legends. Symbols correspond to viscous bedload transport ($s=2.65$, $\mathrm{Ga}=0.1$), turbulent bedload transport ($s=2.65$, $\mathrm{Ga}=20$)  and turbulent saltation transport ($s=2000$, $\mathrm{Ga}=5$). For the bedload transport conditions, the restitution coefficient has been varied to mimic the minimal ($e=0.9$) and nearly maximal ($e=0.01$) possible effect of lubrication forces.}
 \label{BagnoldInterfaceTrends}
\end{figure}

Figures~\ref{BagnoldInterface} and \ref{BagnoldInterfaceTrends} show that the interface $z=z_r$ defined by Eq.~(\ref{InterfaceDef}) obeys Properties~\ref{Property1}-\ref{Property3} of the Bagnold interface for most simulated conditions.
In fact, the numerical data support that most transport ($80{-}100\%$) occurs above $z_r$ [Figs.~\ref{BagnoldInterface}(a) and \ref{BagnoldInterfaceTrends}(a)], that the bed friction coefficient $\mu_b$ does not change much with $\tau$ [Figs.~\ref{BagnoldInterface}(b) and \ref{BagnoldInterfaceTrends}(b)], and that the expression $-P_{zx}(z_r)=\tau-\tau_t$ is approximately obeyed for conditions with $\sqrt{s}\mathrm{Ga}\gtrapprox10$ [Figs.~\ref{BagnoldInterface}(c) and \ref{BagnoldInterfaceTrends}(c)]. Furthermore, $\mu_b$ varies overall between about $0.5$ and $0.9$ with environmental parameters different from $\tau$ [Fig.~\ref{BagnoldInterface}(b)], which is surprisingly small given the large variability of the simulated conditions. That is, $\mu_b$ can be considered an approximate universal constant for the purpose of sediment transport modeling, which is, indeed, what we did in a recent study \cite{PahtzDuran18a}. In contrast, interfaces defined through a constant value of $\phi/\phi_b$ [line-connected symbols in Fig.~\ref{BagnoldInterfaceTrends}(b)], through a constant value of $\mu$ [line-connected symbols in Fig.~\ref{BagnoldInterfaceTrends}(c)], or through other definitions proposed in the literature (not shown) in general do not fulfill the requirements of the Bagnold interface.

Conditions with $\sqrt{s}\mathrm{Ga}\lesssim10$ deviate from Property~\ref{Property3} [Figs.~\ref{BagnoldInterface}(c) and \ref{BagnoldInterfaceTrends}(c)], the reason for which can be seen in Fig.~\ref{BagnoldInterfaceTrends}(d). It shows that the local fluid shear stress $\tau_f=\tau+P_{zx}$ at $z_r$ is near the flow threshold $\tau_t$ at low transport stages and remains constant or decreases with increasing $\Theta$, consistent with Property~\ref{Property3}. However, once a critical value $\Theta\approx0.5$ is exceeded, $\tau_f(z_r)$ begins to increase and enters a regime in which it becomes proportional to $\Theta_t\tau$. This proportionality causes $-P_{zx}(z_r)/\tau$ to approach a limiting value at large transport stages that is smaller than the value unity required by Property~\ref{Property3}, with larger values of the flow threshold Shields number $\Theta_t$ corresponding to larger deviations. In fact, the sediment transport regime that exhibits the largest values of the flow threshold for cohesionless particles [$\max(\Theta_t)\approx0.2$] is viscous bedload transport, which is characterized by comparably small values of $\sqrt{s}\mathrm{Ga}$ \cite{PahtzDuran18a}.

\section{Physical origin of friction law} \label{PhysicalOrigin}
As explained in Sec.~\ref{CoulombFriction}, there have been two interpretations of the friction law (Property~\ref{Property2}) in the literature. In Sec.~\ref{DenseRheologyFailure}, we show that the first interpretation based on the rheology of dense granular flows and suspensions in general is inconsistent with data from our direct transport simulations. In particular, we present strong evidence for the absence of a liquidlike flow regime at low transport stages. In Sec.~\ref{ReboundInterpretation}, we show that the second interpretation associated with particle rebounds at the bed surface is consistent with the simulation data for most conditions. In particular, we explain why this kinematic interpretation also applies to bedload transport, in which the particle dynamics are dominated by long-lasting intergranular contacts rather than particle kinematics.

\subsection{Dense rheology interpretation of friction law} \label{DenseRheologyFailure}
\begin{figure}[htb]
 \begin{center}
  \includegraphics[width=1.0\columnwidth]{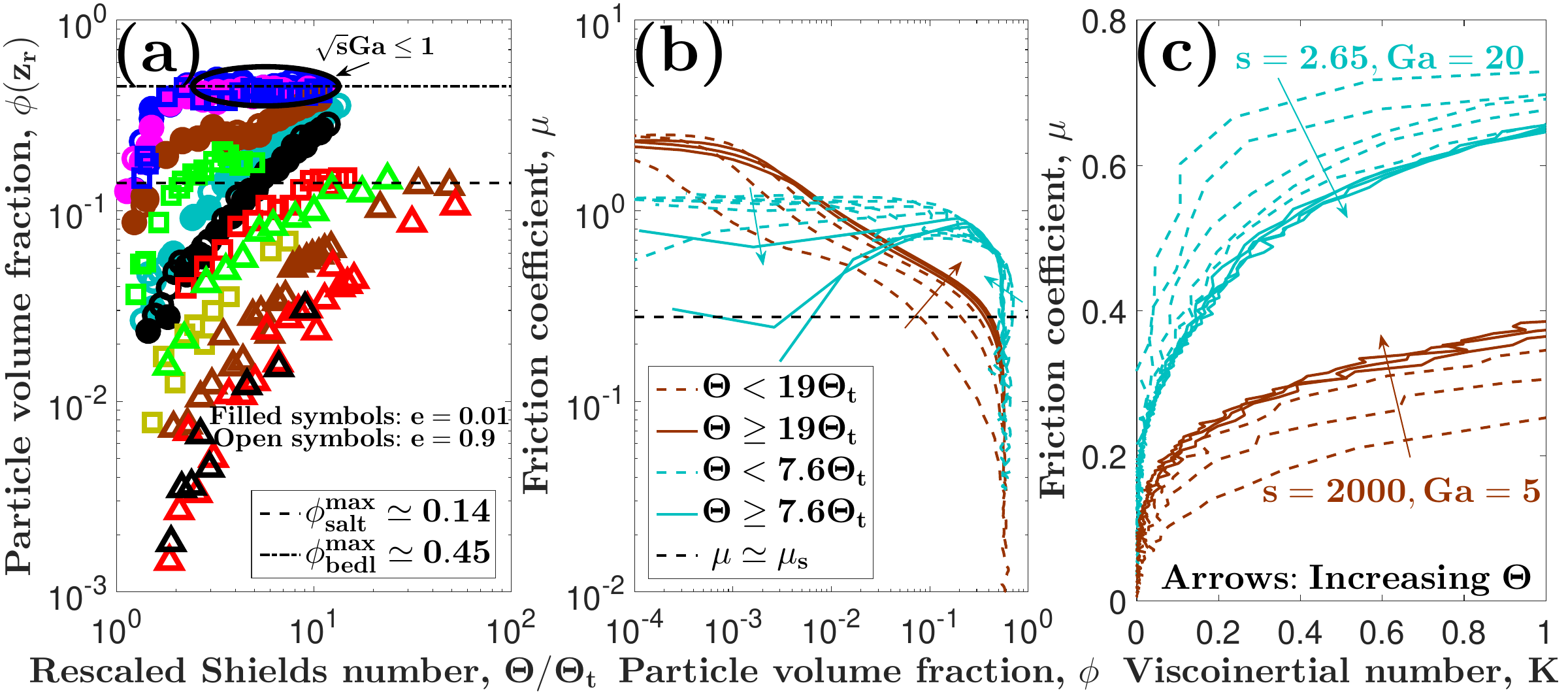}
 \end{center}
 \caption{{\bf{Failure of dense rheology interpretation.}} (a) Particle volume fraction $\phi(z_r)$ at the Bagnold interface versus Shields number $\Theta$. (b, c) Friction coefficient $\mu$ versus (b) particle volume fraction and (c) viscoinertial number $K$. Symbols in (a) correspond to data from our direct transport simulations for various combinations of the particle-fluid-density ratio $s$, Galileo number $\mathrm{Ga}$, and $\Theta$. For symbol legend, see Fig.~\ref{BagnoldInterface}. For conditions with $s\leq2.65$ (corresponding to bedload transport), the restitution coefficient has been varied to mimic the minimal ($e=0.9$) and nearly maximal ($e=0.01$) possible effect of lubrication forces. The turquoise and brown lines in (b) and (c) correspond to the conditions $(s,\mathrm{Ga},e)=(2.65,20,0.9)$ and $(s,\mathrm{Ga},e)=(2000,5,0.9)$, respectively, which are representative for turbulent bedload and saltation transport, respectively.}
 \label{DenseRheologyInterpretation}
\end{figure}

Figure~\ref{DenseRheologyInterpretation}(a) shows that the particle volume fraction $\phi(z_r)$ at the Bagnold interface, obtained from our direct transport simulations, increases with the Shields number $\Theta$ until it approaches at large $\Theta$ a constant maximal value that depends on whether the simulated condition corresponds to bedload ($\phi^{\mathrm{max}}_{\mathrm{bedl}}\simeq0.45$) or saltation transport ($\phi^{\mathrm{max}}_{\mathrm{salt}}\simeq0.14$).
This behavior rules out the dense rheology interpretation of the friction law for most conditions as the liquidlike regime requires $\phi\gtrsim0.4$, particularly when considering that the values of $\phi(z_r)$ are near $10^{-3}$ for some simulated conditions and could possibly be even lower for conditions more extreme than those simulated. However, conditions corresponding to sufficiently intense bedload transport [e.g., conditions with $\sqrt{s}\mathrm{Ga}\leq1$ and $\Theta\gtrsim2\Theta_t$; see ellipse in Fig.~\ref{DenseRheologyInterpretation}(a)] pose a notable exception as $\phi(z_r)\gtrsim0.4$. For these conditions, the dense rheology interpretation of the friction law may, indeed, be consistent with the simulation data.

\subsubsection*{Absence of liquidlike granular flow regime}
The simulation data indicate that a liquidlike granular flow regime does not necessarily exist. For example, Fig.~\ref{DenseRheologyInterpretation}(b) shows for saltation transport with sufficiently low $\Theta/\Theta_t$ (brown, dashed lines) that the local friction coefficient $\mu$ can remain well below the yield stress ratio $\mu_s\simeq0.277$ \cite{Trulssonetal12} within the dense flow region ($\phi\gtrsim0.4$). Furthermore, the thickness of the transient zone in which the particle volume fraction changes from quasistatic ($\phi\simeq0.58$) to gaslike ($\phi\lesssim0.4$) values is, regardless of the transport regime, very thin ($<d$) at low transport stages (Fig.~4 in Ref.~\cite{Duranetal12}, which is based on the same numerical model). In this transient zone and slightly beyond, the average particle velocity $\langle v_x\rangle$ and thus the particle shear rate $\dot\gamma$ obey an exponential relaxation behavior (Fig.~7 in Ref.~\cite{Duranetal12}), and the Bagnold interface ($z=z_r$) is located within this relaxation zone [Fig.~2(a) in Ref.~\cite{PahtzDuran17}, which is based on the same numerical model]. Hence, one may interpret the Bagnold interface as the base of the gaslike transport layer.

Furthermore, an exponential relaxation of $\dot\gamma$ is reminiscent of granular creeping \cite{Bouzidetal13,Bouzidetal15,Houssaisetal15}, which is associated with a nonlocal rheology \cite{Nicholetal10,Reddyetal11,Bouzidetal13,Bouzidetal15}. In fact, if the rheology was local, $\mu$ would solely depend on the particle volume fraction $\phi$ or, alternatively, on the dimensionless number that characterizes the rapidness of the granular shearing motion relative to particle rearrangement processes: the viscoinertial number \cite{Trulssonetal12,NessSun15,NessSun16,Amarsidetal17}
\begin{equation}
 K=\sqrt{(\rho_pd^2\dot\gamma^2+2\rho_f\nu\dot\gamma)/P_{zz}}\equiv\sqrt{I^2+2J}.
\end{equation}
The viscoinertial number $K$ reconciles inertial granular flows, characterized by the inertial number $I=\dot\gamma d/\sqrt{P_{zz}/\rho_p}$, with viscous suspensions, characterized by the viscous number $J=\rho_f\nu\dot\gamma/P_{zz}$. However, a data collapse of $\mu(\phi)$ and $\mu(K)$ is found only when $\Theta$ is sufficiently far from the flow threshold $\Theta_t$ (consistent with Ref.~\cite{Maurinetal16}), where ``sufficiently'' usually refers to relatively intense transport conditions, as shown in Figs.~\ref{DenseRheologyInterpretation}(b) and \ref{DenseRheologyInterpretation}(c) for two cases that are exemplary for turbulent bedload (turquoise lines) and saltation transport (brown lines). 

Put together, the fact that $\mu<\mu_s$ within the dense flow region, the very thin creepinglike transient zone from quasistatic to gaslike particle volume fractions, and the absence of a local and thus liquidlike rheology are strong evidence for a granular solid-gas transition around the Bagnold interface, where the solidlike and gaslike regime are connected by the creepinglike zone. Note that a granular solid-gas transition and the absence of a liquidlike granular flow regime at low transport stages are rather unusual in the context of granular flows and suspensions. To our knowledge, they have previously been found only in viscous bedload transport experiments \cite{Houssaisetal16}. Further note that the absence of a liquidlike rheology at low transport stages implies that two-phase flow models of sediment transport that are based on local rheology models \cite{Chiodietal14,Maurinetal16} can be applied only to sufficiently intense transport conditions.

\subsubsection*{Very viscous bedload transport}
\begin{figure}[htb]
 \begin{center}
  \includegraphics[width=1.0\columnwidth]{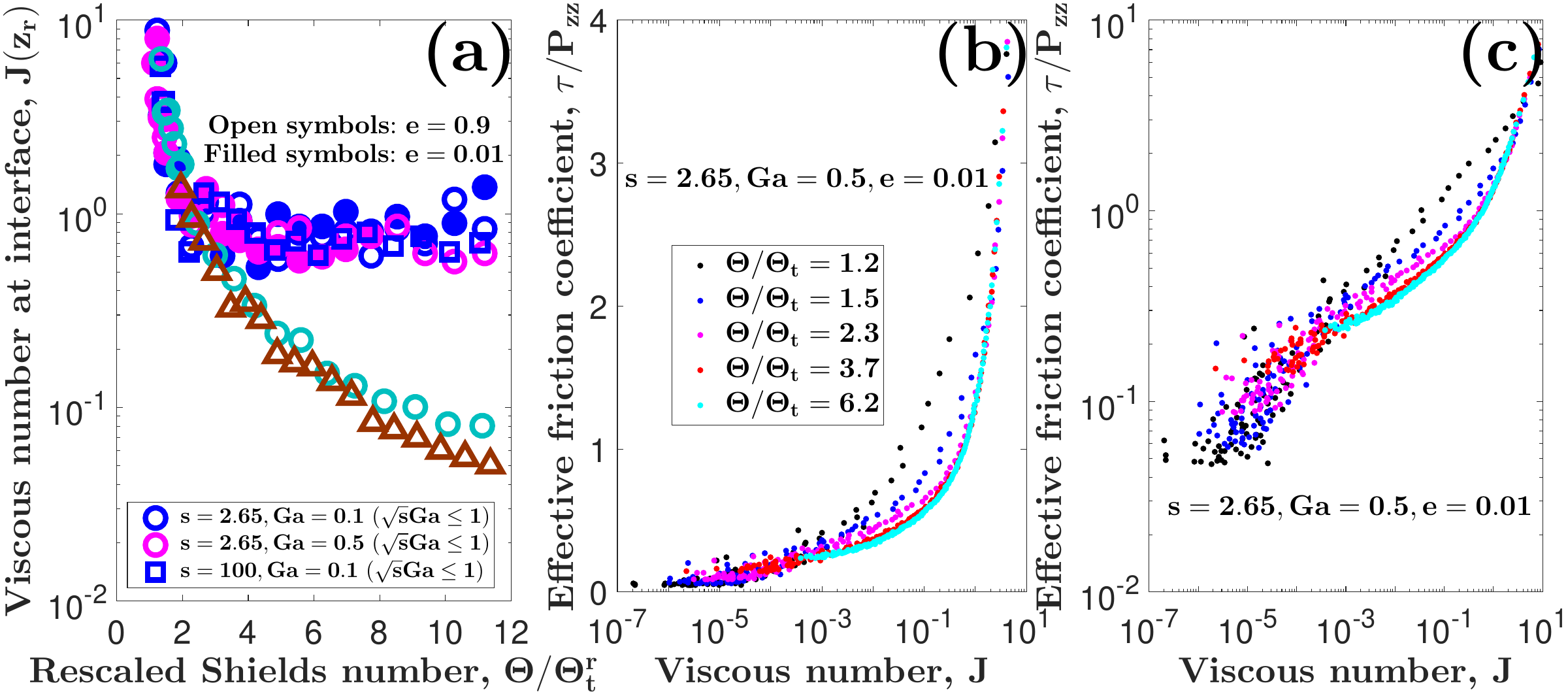}
 \end{center}
 \caption{{\bf{Dense rheology interpretation for very viscous bedload transport.}} (a) Viscous number $J(z_r)$ at the Bagnold interface versus rescaled Shields number $\Theta/\Theta_t$. Symbols correspond to data from our direct transport simulations for those combinations of the particle-fluid-density ratio $s$, Galileo number $\mathrm{Ga}$, and Shields number $\Theta$ that obey $\sqrt{s}\mathrm{Ga}\leq1$. The two conditions $(s,\mathrm{Ga},e)=(2.65,20,0.9)$ (turbulent bedload transport, turquoise circles) and $(s,\mathrm{Ga},e)=(2000,5,0.9)$ (turbulent saltation transport, brown triangles) from Figs.~\ref{DenseRheologyInterpretation}(b) and \ref{DenseRheologyInterpretation}(c) are also shown for comparison. (b, c) Effective friction coefficient $\tau/P_{zz}$ versus $J$ for the case $(s,\mathrm{Ga},e)=(2.65,0.5,0.01)$ and several $\Theta/\Theta_t$ in (b) log-linear and (c) log-log scale.}
 \label{ViscousBedload}
\end{figure}

For conditions corresponding to very viscous bedload transport ($\sqrt{s}\mathrm{Ga}\leq1$), the absence of a liquidlike granular flow regime is limited to Shields numbers relatively close to the flow threshold ($\Theta\lesssim2\Theta_t$). In fact, for $\Theta\gtrsim2\Theta_t$, both the friction coefficient $\mu$ [Figs.~\ref{BagnoldInterface}(b) and \ref{BagnoldInterfaceTrends}(b)] and particle volume fraction $\phi$ [ellipse in Fig.~\ref{DenseRheologyInterpretation}(a)] are approximately constant at $z_r$, which is consistent with a local rheology $\mu(\phi)$ around the Bagnold interface (i.e., liquidlike flow behavior due to $\phi\gtrsim0.4$). Figure~\ref{ViscousBedload}(a) shows that very viscous bedload transport conditions (but no other conditions) also exhibit an approximately constant value of the viscous number $J(z_r)$ for $\Theta\gtrsim2\Theta_t$, which is consistent with a local rheology $\mu(J)$.
Consistently, Figs.~\ref{ViscousBedload}(b) and \ref{ViscousBedload}(c) show exemplary for the case $(s,\mathrm{Ga},e)=(2.65,0.5,0.01)$ that the simulation data of the effective friction coefficient $\tau/P_{zz}$ collapse as a function of $J$ for sufficiently large $\Theta/\Theta_t$, whereas this local rheology behavior is disobeyed for small $\Theta/\Theta_t$. This finding and the shape of the profiles of $[\tau/P_{zz}](J)$ shown in Figs.~\ref{ViscousBedload}(b) and \ref{ViscousBedload}(c) are in qualitative agreement with recent viscous bedload transport measurements (cf. Fig.~9 in Ref.~\cite{Houssaisetal16}).

We now show that the approximate constancy of $J(z_r)$ for sufficiently large $\Theta/\Theta_t$ can be inferred from the definition of the Bagnold interface [Eq.~(\ref{InterfaceDef})] applied to viscous conditions. First, using $\mu=-P_{zx}/P_{zz}$ and the fact that the local viscous fluid shear stress can be expressed as $\tau_f=\tau+P_{zx}=\rho_f\nu(1-\phi)u_x^\prime$ \cite{Duranetal12,PahtzDuran17}, where $u_x$ is the mean horizontal fluid velocity, we obtain from Eq.~(\ref{InterfaceDef}) that the following condition must be obeyed at the Bagnold interface ($z=z_r$):
\begin{equation}
 (P_{zz}\dot\gamma)^\prime=P_{zz}\dot\gamma^\prime-\mu^\prime P_{zx}\dot\gamma-\rho_f\nu\mu\dot\gamma[(1-\phi)u_x^\prime]^\prime=0. \label{ViscousCondition1}
\end{equation}
Second, we neglect spatial changes of the particle volume fraction $\phi$ because it is close to the packing fraction in dense systems, and thus we also neglect spatial changes of $\mu$ as they are of the same order \cite{Trulssonetal12}. Using these approximations and the shear rate definition $\dot\gamma=\langle v_x\rangle^\prime$ in Eq.~(\ref{ViscousCondition1}), we approximately obtain
\begin{equation}
 J(z_r)\approx\frac{[\langle v_x\rangle^{\prime\prime}/u_x^{\prime\prime}](z_r)}{\mu_b[1-\phi(z_r)]}.
\end{equation}
The quantity $[\langle v_x\rangle^{\prime\prime}/u_x^{\prime\prime}](z_r)$ is expected to exhibit an approximately constant value smaller than unity as the particle velocity profile $\langle v_x\rangle(z)$ is strongly coupled to the flow velocity profile $u_x(z)$ when the bed is fully mobile (i.e., liquidlike) due to a strong viscous drag forcing \cite{PahtzDuran17}, which explains the approximate constancy of $J(z_r)$ for sufficiently large $\Theta/\Theta_t$ (Fig.~\ref{ViscousBedload}a). Hence, for conditions corresponding to very viscous bedload transport ($\sqrt{s}\mathrm{Ga}\leq1$) sufficiently far from the flow threshold ($\Theta\gtrsim2\Theta_t$), $\mu_b\approx\mathrm{const}$ can be explained in the context of dense granular flows and suspensions.

\subsection{Rebound interpretation of friction law} \label{ReboundInterpretation}
\begin{figure}[htb]
 \begin{center}
  \includegraphics[width=0.5\columnwidth]{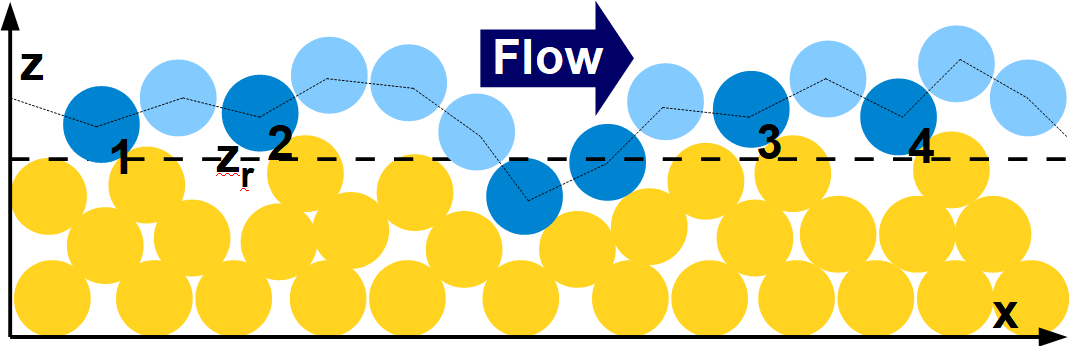}
 \end{center}
 \caption{{\bf{Sketch of the trajectory of a particle hopping along a granular bed.}} Driven by the flow, a transported particle (blue) hops along the solidlike or liquidlike granular bed (yellow particles). Instants of particle contacts are colored deep blue, and the ones for which the center of mass of the transported particle is above the Bagnold interface ($z>z_r$) are numbered consecutively (for illustrating the mathematical derivation in the Appendix).}
 \label{ReboundSketch}
\end{figure}

The gaslike transport layer is composed of particles that hop, slide, and/or roll along a solidlike granular bed at low transport stages or a liquidlike granular bed at large transport stages [Figs.~\ref{DenseRheologyInterpretation}(b) and \ref{DenseRheologyInterpretation}(c)]. Except for very viscous bedload transport (which is therefore excluded from the following considerations), the hopping motion is significant and usually even dominates above the Bagnold interface ($z>z_r$) \cite{PahtzDuran18a}. Now we argue that a steady transport state in which particles hop along a granular bed (Fig.~\ref{ReboundSketch}) causes the kinetic friction coefficient $\mu^t\equiv-P^t_{zx}/P^t_{zz}$ to be approximately constant at $z_r$: $\mu^t_b\equiv\mu^t(z_r)\approx\mathrm{const}$.

\subsubsection*{Constant kinetic friction coefficient}
First, defining the average $\langle A\rangle_{\uparrow(\downarrow)}=\phi\langle AH[+(-)v_z]\rangle/\phi_{\uparrow(\downarrow)}$ of a quantity $A$ over ascending (descending) particles, where $H$ the Heaviside function and $\phi_{\uparrow(\downarrow)}=\phi\langle H[+(-)v_z]\rangle$ the volume fraction of ascending (descending) particles, we approximately obtain
\begin{equation}
 \phi\langle v_zv_i\rangle=\phi_\uparrow\langle v_zv_i\rangle_\uparrow+\phi_\downarrow\langle v_zv_i\rangle_\downarrow\approx\phi_\uparrow\langle v_z\rangle_\uparrow\langle v_i\rangle_\uparrow+ \phi_\downarrow\langle v_z\rangle_\downarrow\langle v_i\rangle_\downarrow=\phi_\uparrow\langle v_z\rangle_\uparrow(\langle v_i\rangle_\uparrow-\langle v_i\rangle_\downarrow), \label{CorrelationApproximation}
\end{equation}
where we neglected velocity correlations and used the steady-state mass balance $\phi\langle v_z\rangle=\phi_\uparrow\langle v_z\rangle_\uparrow+\phi_\downarrow\langle v_z\rangle_\downarrow=0$ \cite{Pahtzetal15a}. Further using the definition of the kinetic stresses [Eq.~(\ref{Pt})] and $\langle c_zc_i\rangle=\langle v_zv_i\rangle$ (which follows from $\langle v_z\rangle=0$), we then obtain from Eq.~(\ref{CorrelationApproximation})
\begin{equation}
 \mu^t=-\frac{P^t_{zx}}{P^t_{zz}}=-\frac{\phi\langle c_zc_x\rangle}{\phi\langle c_z^2\rangle}=-\frac{\phi\langle v_zv_x\rangle}{\phi\langle v_z^2\rangle}\approx\frac{\langle v_x\rangle_\downarrow-\langle v_x\rangle_\uparrow}{\langle v_z\rangle_\uparrow-\langle v_z\rangle_\downarrow}. \label{mutapprox}
\end{equation}
As the Bagnold interface is the effective elevation of energetic particles rebounding at the bed surface (Sec.~\ref{Interface}), Eq.~(\ref{mutapprox}) implies that $\mu^t_b$ is a measure for the ratio between the average horizontal momentum loss [$\propto(\langle v_x\rangle_\downarrow-\langle v_x\rangle_\uparrow)(z_r)$] and vertical momentum gain [$\propto(\langle v_z\rangle_\uparrow-\langle v_z\rangle_\downarrow)(z_r)$] of hopping particles rebounding at the bed surface.

Second, provided that the influence of fluid drag on the vertical motion of hopping particles can be neglected (this precondition is indirectly verified by the fact that the final result is consistent with data from our direct transport simulations), a steady hopping motion requires $\langle v_z\rangle_\uparrow(z_r)\approx-\langle v_z\rangle_\downarrow(z_r)$ due to energy conservation. On average, only an approximately constant impact angle $\alpha_i=-\arctan[\langle v_z\rangle_\downarrow/\langle v_x\rangle_\downarrow](z_r)$, resulting in an approximately constant rebound angle $\alpha_r=\arctan[\langle v_z\rangle_\uparrow/\langle v_x\rangle_\uparrow](z_r)$, can ensure this constraint \cite{JenkinsValance14,Berzietal16,Berzietal17}, which combined implies $\mu^t_b\approx\mathrm{const}$.

\subsubsection*{Approximate equality of friction coefficients}
\begin{figure}[htb]
 \begin{center}
  \includegraphics[width=1.0\columnwidth]{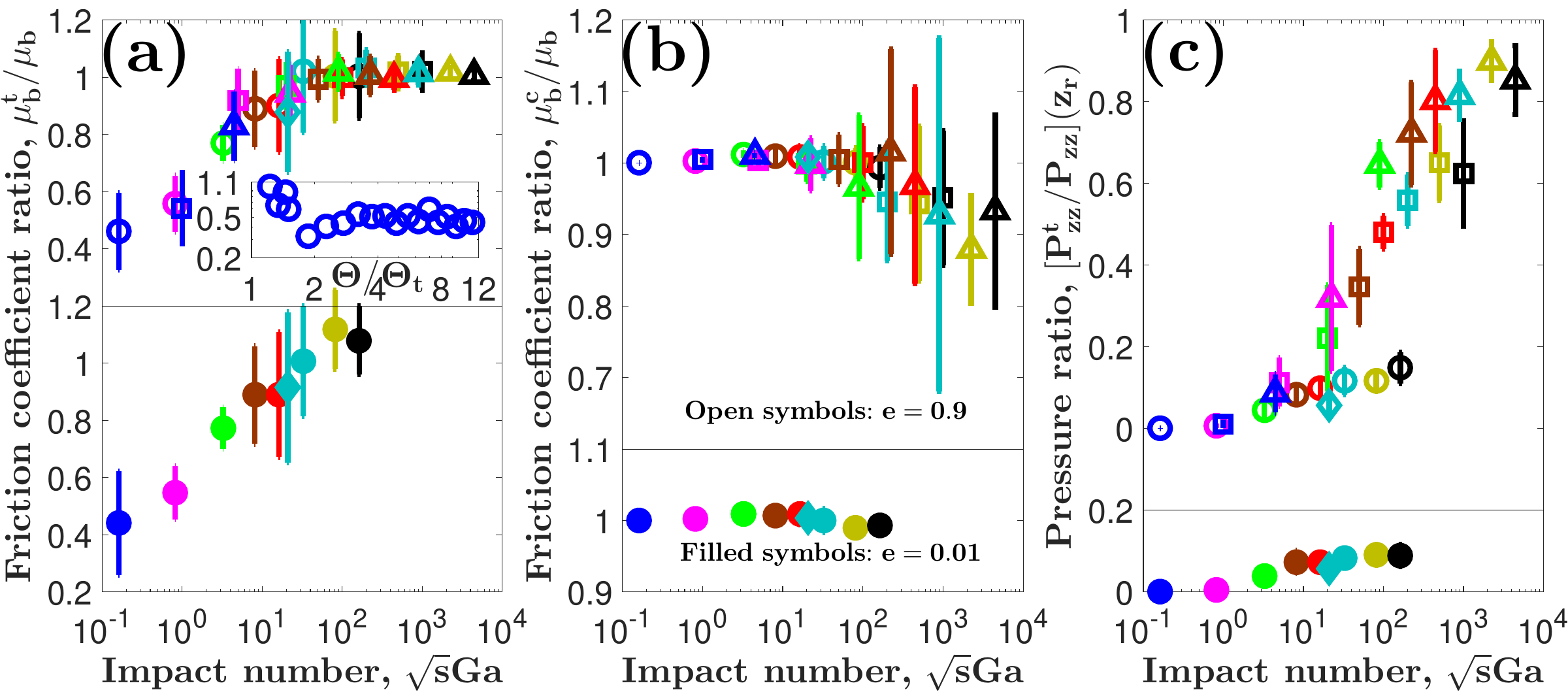}
 \end{center}
 \caption{{\bf{Approximate equality of friction coefficients.}} Friction ratios (a) $\mu^t_b/\mu_b$ and (b) $\mu^c_b/\mu_b$ and pressure ratio (c) $[P^t_{zz}/P_{zz}](z_r)$ versus impact number $\sqrt{s}\mathrm{Ga}$. Symbols correspond to data from our direct transport simulations for various combinations of the particle-fluid-density ratio $s$, Galileo number $\mathrm{Ga}$, and Shields number $\Theta$. For symbol legend, see Fig.~\ref{BagnoldInterface}. For conditions with $s\leq2.65$ (corresponding to bedload transport), the restitution coefficient has been varied to mimic the minimal ($e=0.9$) and nearly maximal ($e=0.01$) possible effect of lubrication forces. The vertical bars indicate the range of values the quantities cover with varying $\Theta$ above about $2\Theta_t$. This lower limit is imposed to separate the random variability due to bad statistics when $\Theta$ is close to $\Theta_t$ [e.g., see Fig.~\ref{BagnoldInterfaceTrends}(c)] from the actual variability. Inset of (a): friction ratio $\mu^t_b/\mu_b$ versus rescaled Shields number $\Theta/\Theta_t$ for very viscous bedload transport ($s=2.65$, $\mathrm{Ga}=0.1$, $e=0.9$).}
 \label{FrictionCoefficients}
\end{figure}

Until here our reasoning is largely in line with previous studies \cite{Sauermannetal01,DuranHerrmann06,Pahtzetal12,Lammeletal12,JenkinsValance14,Berzietal16,Berzietal17}. These studies now concluded $\mu_b\approx\mathrm{const}$ from $\mu^t_b\approx\mathrm{const}$, which is consistent with our direct transport simulations, as shown in Fig.~\ref{FrictionCoefficients}(a).
In fact, it can be seen that $\mu^t_b/\mu_b$ is relatively close unity for most simulated conditions, except for very viscous bedload transport conditions ($\sqrt{s}\mathrm{Ga}\leq1$) with $\Theta/\Theta_t\gtrsim2$. However, exactly for these conditions, $\mu_b\approx\mathrm{const}$ has been explained from the local rheology of dense viscous suspension (Sec.~\ref{DenseRheologyFailure}). Interestingly, conditions with $\sqrt{s}\mathrm{Ga}\leq1$ and $\Theta/\Theta_t\lesssim2$ exhibit values of $\mu^t_b/\mu_b$ that are again relatively close to unity, as shown for an exemplary case in the inset of Fig.~\ref{FrictionCoefficients}(a). This suggests that the rebound interpretation of $\mu_b\approx\mathrm{const}$ explained in this section may actually apply to very viscous bedload transport at low transport stages even though the hopping motion is dominated by particles sliding and rolling along the granular bed \cite{PahtzDuran18a}.

Figure~\ref{FrictionCoefficients}(b) shows that the contact friction coefficient $\mu^c\equiv-P^c_{zx}/P^c_{zz}$ is relatively close to $\mu_b$ for all simulated conditions. Furthermore, Fig.~\ref{FrictionCoefficients}(c) tests the hypothesis of previous studies \cite{Sauermannetal01,DuranHerrmann06,Pahtzetal12,Lammeletal12,JenkinsValance14,Berzietal16,Berzietal17} that $P^t_{ij}(z_r)\approx P_{ij}(z_r)$ is the reason why $\mu^t_b\approx\mu_b$. It can be seen that, while this reasoning works well for saltation transport conditions, it does not hold for bedload transport conditions because $\mu^t_b\approx\mu_b$ despite $P^t_{ij}(z_r)\ll P_{ij}(z_r)$.

In the Appendix, we derive $\mu^t_b\approx\mu^c_b\approx\mu_b$ from first physical principles. In summary, this derivation mainly exploits that the granular transport layer is gaslike, which means that collisions between particles located above the Bagnold interface are predominantly binary. This property allows us to write the contact stress tensor component $P^c_{zi}(z_r)$ as the total impulse per unit bed area per unit time generated by collisions between particles transported above the Bagnold interface with bed particles below the Bagnold interface. Using Eq.~(\ref{CorrelationApproximation}) (which is based on the steady-state mass balance) and that the Bagnold interface ($z=z_r$) is the effective elevation of energetic particle-bed rebounds (Sec.~\ref{Interface}), it can then be shown that each such collision approximately generates the impulse equivalent per unit bed area per unit time of the associated kinetic stress tensor component $P^t_{zi}(z_r)$, which implies $P^c_{zi}(z_r)\approx\overline{R}_{\uparrow z_r}P^t_{zi}(z_r)$, where $\overline{R}_{\uparrow z_r}$ is the average number of such collisions per crossing of the Bagnold interface from below. As $\overline{R}_{\uparrow z_r}$ is the same for $i=x$ and $i=z$, it eventually follows $\mu^c_b\approx\mu^t_b$ and thus $\mu^t_b\approx\mu^c_b\approx\mu_b$.

\section{Discussion and Conclusions} \label{Conclusions}
We have used numerical simulations that couple the discrete element method for the particle motion with a continuum Reynolds-averaged description of hydrodynamics to study the physical origin and universality of theoretical threshold shear stress-based models of the rate of nonsuspended sediment transport for a large range of Newtonian fluids driving transport, including viscous and turbulent liquids and air. The vast majority of such models are based on, or can be reformulated in the spirit of, Bagnold's \cite{Bagnold56,Bagnold66,Bagnold73} assumption that there is a well-defined interface between granular bed and transport layer, which we have called the ``Bagnold interface'', with certain special properties (Properties~\ref{Property1}-\ref{Property3} in the Introduction). From our study, we have gained the following insights:
\begin{enumerate}
 \item Our simulations support the hypothesis that the Bagnold interface corresponds to the effective elevation at which the most energetic particles rebound, which can be mathematically defined through a maximum of the local production rate of cross-correlation fluctuation energy [Eq.~(\ref{InterfaceDef})].
 \item Our simulations indicate that, in general, the transition between the solidlike granular bed and gaslike granular transport layer occurs through a very thin granular creepinglike zone, which contains the Bagnold interface and which is associated with a nonlocal granular flow rheology. A local rheology, which is required for liquidlike behavior, is usually found only for relatively intense transport conditions [Figs.~\ref{DenseRheologyInterpretation}(b) and \ref{DenseRheologyInterpretation}(c)]. The absence of a liquidlike rheology at low transport stages implies that two-phase flow models of sediment transport that are based on local rheology models \cite{Chiodietal14,Maurinetal16} can be applied only to sufficiently intense transport conditions.
 \item As the majority of sediment transport is gaslike, the transport rate above the Bagnold interface well approximates the overall transport rate, as supported by our simulations [Figs.~\ref{BagnoldInterface}(a) and \ref{BagnoldInterfaceTrends}(a)] and demanded by Property~\ref{Property1}.
 \item Our simulations indicate that the ratio between the particle shear stress and normal-bed pressure at the Bagnold interface, the bed friction coefficient $\mu_b$, varies between about $0.5$ and $0.9$ for the entire range of simulated conditions [Figs.~\ref{BagnoldInterface}(b) and \ref{BagnoldInterfaceTrends}(b)]. In particular, $\mu_b$ is insensitive to the fluid shear stress $\tau$, as demanded by Property~\ref{Property2}. The physical origin of this universal approximate invariance of $\mu_b$ has been physically linked to a steady transport state in which particles continuously rebound at the bed surface (Fig.~\ref{ReboundSketch}).
 \item Very viscous bedload transport ($\sqrt{s}\mathrm{Ga}\lesssim1$) not too far above the flow threshold ($\Theta\gtrsim2\Theta_t$) poses a notable exception: our simulations indicate that the granular flow around the Bagnold interface is liquidlike [Figs.~\ref{DenseRheologyInterpretation}(a) and \ref{ViscousBedload}], and the friction law has been physically linked to the local rheology of dense viscous suspensions.
 \item As the friction law is obeyed at the base of the gaslike transport layer, $\mu_b$ fundamentally differs from the constant yield stress ratio associated with the solid-liquid transition in dense granular flows and suspensions. This finding challenges a large number of studies \cite{Bagnold56,Bagnold66,Bagnold73,AshidaMichiue72,EngelundFredsoe76,KovacsParker94,NinoGarcia94,NinoGarcia98a,Seminaraetal02,Parkeretal03,AbrahamsGao06} according to which $\mu_b$ is the yield stress ratio. 
 \item Our simulations indicate that the local fluid shear stress $\tau_f(z_r)$ at the Bagnold interface reduces to a value near the flow threshold $\tau_t$ at low transport stages and remains constant or decreases with increasing Shields number $\Theta$, consistent with Property~\ref{Property3}. However, once a critical value $\Theta\approx0.5$ is exceeded, $\tau_f(z_r)$ begins to increase and enters a regime in which it becomes proportional to $\Theta_t\tau$. This behavior results in a deviation from Property~\ref{Property3} for sufficiently viscous bedload transport ($\sqrt{s}\mathrm{Ga}\lesssim10$).
\end{enumerate}

Concerning the last point, it is commonly argued that $\tau_f(z_r)$ reduces to the smallest value that just allows entrainment of bed sediment (by the splash caused by particle-bed impacts and/or by the action of fluid forces), which is assumed to be near $\tau_t$ \cite{Bagnold56,Bagnold66,Bagnold73,AshidaMichiue72,EngelundFredsoe76,KovacsParker94,NinoGarcia94,NinoGarcia98a,Seminaraetal02,Parkeretal03,AbrahamsGao06,Kawamura51,Owen64,Kind76,LettauLettau78,Sorensen91,Sorensen04,Sauermannetal01,DuranHerrmann06,Pahtzetal12,Lammeletal12}. However, according to our recent study \cite{PahtzDuran18a}, $\tau_t$ is not an entrainment threshold but rather a rebound threshold: the minimal fluid shear stress needed to compensate the average energy loss of rebounding particles by the fluid drag acceleration during particle trajectories. That is, $\tau_f(z_r)$ reduces to the smallest value that just allows a long-lasting rebound motion. This interpretation (which was originally proposed by Bagnold \cite{Bagnold41} for turbulent saltation transport but later discarded) is independent of whether the bed is rigid or erodible and consistent with our finding that $\mu_b\approx\mathrm{const}$ is linked to a steady rebound state rather than the constant yield stress ratio at the granular solid-liquid transition. In fact, based on this rebound picture, we have proposed a universal analytical flow threshold model \cite{PahtzDuran18a}, which uses $\mu_b=0.63$ (the simulation mean) and which predicts $\tau_t$ for arbitrary environmental conditions in simultaneous agreement with available measurements in air and viscous and turbulent liquids despite not being fitted to any kind of experimental data. That is, the only ingredient that remains missing for a universal scaling law predicting the rate of nonsuspended sediment transport [i.e., a version of Eq.~(\ref{Q}) that is applicable to arbitrary environmental conditions] is a universal scaling law for the average particle velocity $\overline{v_x}$ in the flow direction. So far, we have succeeded in deriving an expression for $\overline{v_x}$ for sufficiently low $\Theta/\Theta_t$ \cite{PahtzDuran18a}, and we are currently working on a generalization to arbitrarily large $\Theta/\Theta_t$. Finally, we would like to emphasize that bed sediment entrainment, even though it does not seem to affect the functional structure of the scaling laws of nonsuspended sediment transport, is still required to sustain the equilibrium state described by such laws \cite{PahtzDuran18a}.

\begin{acknowledgments}
We acknowledge support from a grant from the National Natural Science Foundation of China (No.~11750410687).
\end{acknowledgments}

\appendix*
\section{Physical derivation of equality of friction coefficients}
First, we use the steady momentum balance with respect to contact forces: $-P^{c\prime}_{zi}=\rho_p\phi\langle a^c_i\rangle$ \cite{Pahtzetal15a}, where $\mathbf{a}^c$ is the particle acceleration due to contact forces ($\mathbf{F}^{cm}=\sum_n\mathbf{F}^{mn}$). Integrating this balance over elevations $z>z_r$ yields
\begin{equation}
 P^c_{zi}(z_r)=\frac{1}{T\Delta}\sum_n\int\limits_0^TF^{cn}_iH(z^n-z_r)\d t, \label{ContactTensor1}
\end{equation}
where we used $\int_{z_r}^\infty\delta(z-z^n)\d z=H(z^n-z_r)$ and Eq.~(\ref{LocalAverage}). Above the Bagnold interface ($z>z_r$), the granular flow is gaslike [Fig.~\ref{DenseRheologyInterpretation}(a)], implying that particle contacts between hopping particles mainly occur during binary collisions. Because a binary contact between a particle $m$ and a particle $n$ does not contribute to Eq.~(\ref{ContactTensor1}) due to $\mathbf{F}^{cm}+\mathbf{F}^{cn}=0$, the contacts contributing to Eq.~(\ref{ContactTensor1}) are predominantly particle-bed rebounds (colored deep blue in Fig.~\ref{ReboundSketch}). The term $\int_0^TF^{cn}_iH(z^n-z_r)\d t$ thus describes the total impulse gained by particle $n$ in time $T$ during those particle-bed rebounds in which its center of mass is located above the Bagnold interface ($z^n>z_r$). Consecutively numbering such particle-bed rebounds by $r^n=1,2,...,R^n_T$ (Fig.~\ref{ReboundSketch}), where $R^n_T$ is the total number of rebounds of particle $n$ that occur in time $T$ above $z_r$, and denoting the velocity change caused by each rebound as $\delta v^{r^n}_i$, which implies that $\rho_pV_p^n\delta v^{r^n}_i$ is the gained impulse at each rebound, we obtain from Eq.~(\ref{ContactTensor1})
\begin{equation}
 P^c_{zi}(z_r)\simeq\frac{1}{T\Delta}\sum_n\sum_{r^n=1}^{R^n_T}\rho_pV_p^n\delta v^{r^n}_i=\frac{\rho_p\overline{\delta v^r_i}}{T\Delta}\sum_n R^n_TV_p^n, \label{ContactTensor2}
\end{equation}
where $\overline{\delta v^r_i}$ is the average of $\delta v^{r^n}_i$ over all particles and particle-bed rebounds above $z_r$. Now we separate $R^n_T$ into the number of instants $\#^{n,T}_{\uparrow z_r}$ particle $n$ crosses the Bagnold interface from below in time $T$ and the average number $R^n_{\uparrow z_r}$ of rebounds of particle $n$ per such crossing that occur above $z_r$: $R^n_T=R^n_{\uparrow z_r}\#^{n,T}_{\uparrow z_r}$. Furthermore, as the Bagnold interface is the effective elevation of energetic particle-bed rebounds (Sec.~\ref{Interface}), we approximate $\delta v^{r^n}_i$ by the average velocity gain at $z_r$: $\delta v^{r^n}_i\approx\langle v_z\rangle_\uparrow(z_r)-\langle v_z\rangle_\downarrow(z_r)$. Combining these mathematical manipulations and using Eqs.~(\ref{Pt}) and (\ref{CorrelationApproximation}), and the fact that the vertical upward-flux $[\phi_\uparrow\langle v_z\rangle_\uparrow](z_r)$ of particles through the Bagnold interface equals the total particle volume $\sum_n\#^{n,T}_{\uparrow z_r}V^n$ that crosses the Bagnold interface from below per unit bed area $\Delta$ per unit time $T$, we approximately obtain from Eq.~(\ref{ContactTensor2})
\begin{equation}
 P^c_{zi}(z_r)\approx\frac{\rho_p[\langle v_z\rangle_\uparrow-\langle v_z\rangle_\downarrow](z_r)}{T\Delta}\sum_n R^n_{\uparrow z_r}\#^{n,T}_{\uparrow z_r}V_p^n=\overline{R}_{\uparrow z_r}\rho_p[\phi_\uparrow\langle v_z\rangle_\uparrow](z_r)[\langle v_z\rangle_\uparrow-\langle v_z\rangle_\downarrow](z_r)\approx\overline{R}_{\uparrow z_r}P^t_{zi}(z_r), \label{ContactTensor3}
\end{equation}
where $\overline{R}_{\uparrow z_r}$ is the average number of particle-bed rebounds above $z_r$ per crossing of the Bagnold interface from below. Equation~(\ref{ContactTensor3}) means that the contact contribution $P^c_{zi}(z_r)$ to the stress tensor $P_{zi}(z_r)$ is approximately proportional to the kinetic contribution $P^t_{zi}(z_r)$, where the proportionality factor $\overline{R}_{\uparrow z_r}$ is the same for $i=x$ and $i=z$. Hence, Eq.~(\ref{ContactTensor3}) implies $\mu^t_b\approx\mu^c_b\approx\mu_b$.

%

\end{document}